\documentclass[fleqn,usenatbib]{mnras}

\usepackage{newtxtext,newtxmath}

\usepackage[T1]{fontenc}

\DeclareRobustCommand{\VAN}[3]{#2}
\let\VANthebibliography\thebibliography
\def\thebibliography{\DeclareRobustCommand{\VAN}[3]{##3}\VANthebibliography}

\usepackage{graphicx}	
\usepackage{amsmath}	
\usepackage{xcolor}
\usepackage{ulem}

\title[Role of the GR in the dynamics of HZ particles]{Effects of general relativity on habitable zone particles under the presence of an inner perturber around solar-mass stars}

\author[C. F. Coronel]
{C. F. Coronel$^{1,2}$%
\thanks{E-mail: ccoronel@fcaglp.unlp.edu.ar},
G. C. de El\'ia$^{1,2}$, 
M. Zanardi$^{1,2}$
and A. Dugaro$^{1,2}$ 
\\
$^{1}$Instituto de Astrof\'{\i}sica de La Plata, CCT La Plata-CONICET-UNLP, Paseo del Bosque S/N (1900), La Plata, Argentina\\ 
$^{2}$Facultad de Ciencias Astron\'omicas y Geof\'{\i}sicas, Universidad Nacional de La Plata, Paseo del Bosque S/N (1900), La Plata, Argentina
}

\date{Accepted XXX. Received YYY; in original form ZZZ}

\pubyear{2023}

\begin{document}
\label{firstpage}
\pagerange{\pageref{firstpage}--\pageref{lastpage}}
\maketitle

\begin{abstract}
We analyze the role of the general relativity (GR) on the nodal librations of test particles located at the Habitable Zone (HZ) around a solar-mass star, which evolve under the influence of an eccentric planetary-mass perturber with a semimajor axis of 0.1 au. Based on a secular Hamiltonian up to quadrupole level, we derive analytical criteria that define the nodal libration region of a HZ particle as a function of its eccentricity $e_2$ and inclination $i_2$, and the mass $m_1$ and the eccentricity $e_1$ of the perturber. We show that a HZ particle can experience nodal librations with orbital flips or purely retrograde orbits for any $m_1$ and $e_1$ by adopting a suitable combination of $e_2$ and $i_2$. For $m_1 <$ 0.84 M$_\textrm{Jup}$, the greater the $m_1$ value, the smaller the $e_2$ value above which nodal librations are possible for a given $e_1$. For $m_1 >$ 0.84 M$_{\textrm{Jup}}$, a HZ test particle can undergo nodal librations for any $e_2$ and appropriate values of $e_1$ and $i_2$. The same correlation between $m_1$ and $e_2$ is obtained for nodal librations with orbital flips, but a mass limit for $m_1$ of 1.68 M$_{\textrm{Jup}}$ is required in this case. Moreover, the more massive the inner perturber, the greater the nodal libration region associated with orbital flips in the ($e_1$, $i_2$) plane for a given value of $e_2$. Finally, we find good agreements between the analytical criteria and results from N-body simulations for values of $m_1$ ranging from Saturn-like planets to super-Jupiters.  
\end{abstract}




\section{Introduction}

The secular dynamics of test particles in the framework of the elliptical restricted three-body problem has been the focus of study of a large number of works in the literature. These investigations were aimed at improving our understanding of several astrophysical phenomena linked to different areas of astronomy. Historically, most such studies focused on the dynamical evolution of an inner test particle orbiting a central star under the influence of a far-away perturber \citep[e.g.][]{vonZeipel1910, Lidov1962, Kozai1962, Lithwick2011, Katz2011, Naoz2016}. Here, we are interested in deepening our understanding of the inverse problem, in which an outer test particle secularly evolves under the effects of an inner perturber around a given star.

A pioneer work concerning the elliptical restricted three-body problem for an outer test particle is that developed by \citet{Ziglin1975}. In this study, the author focused on the analysis of the secular evolution of an outer planet of negligible mass orbiting a binary-star system. To do this, \citet{Ziglin1975} studied an integrable limiting case of the doubly averaged disturbing function of the elliptical restricted three-body problem. From this, the author showed that a circular binary only leads to nodal circulations of the outer test particle, while the greater the binary's eccentricity, the wider the range of inclinations associated with the nodal libration region.

During the last fifteen years, the elliptical restricted three-body problem for an outer test particle has received much attention by various authors. In this line of research, \citet{Verrier2009} investigated the problem through numerical and analytical models, obtaining an empirical criteria for the high-inclination stability limits in general triple systems. Then, \citet{FaragoLaskar2010} studied the case of a distant body orbiting an inner binary in the secular and quadrupolar approximations. These authors derived results consistent with those obtained by \citet{Ziglin1975} and extended their research to the general three-body problem. Later, \citet{Gallardo2012} analyzed the inverse Lidov-Kozai resonance for trans-Neptunian objects considering the gravitational perturbations of the giant planets assumed on circular and coplanar orbits. After that, \citet{Li2014} and \citet{Naoz2017} obtained analytical solutions to some orbital elements of circumbinary orbits from a quadrupole secular theory and explored the role of the octupole level of the secular Hamiltonian. Moreover, \citet{Naoz2017} briefly discussed the effects of the general relativity (GR) in the dynamics of the system. Then, \citet{Vinson2018} analyzed secular resonances in the outer restricted three-body problem from a Hamiltonian expanded to hexadecapole level. On the basis on this approximation, \citet{deElia2019} studied the inverse Lidov-Kozai resonance for an outer test particle around a binary for a wide range of orbital parameters. Later, \citet{Hansen2020} analyzed the stationary points of the hierarchical three-body problem at both the quadrupole and octupole levels.

\citet{Zanardi2018} developed a significant contribution to this line of research, studying in detail the role of the GR in the elliptical restricted three-body problem for an outer test particle. These authors derived general analytical criteria for nodal librations of circumbinary test particles, which strongly depend on the physical and orbital properties of the bodies of the system. By making use of the prescriptions obtained by \citet{Zanardi2018}, \citet{Lepp2022ApJ} found a radial limit to nodal librations of outer test particles on circular orbits around a binary-star system from GR effects. Simultaneously to the present research, \citet{Zanardi2023} refined the criteria derived by \citet{Zanardi2018} and obtained constraints to the semimajor axis of outer particles with nodal librations in the elliptical restricted three-body problem by GR effects. These authors considered an inner binary composed of a star and a planetary-mass companion and analyzed the sensitivity of the results to the mass of the star, the mass, the semimajor axis and the eccentricity of the inner planetary-mass perturber, and the eccentricity and the inclination of the outer test particle.   

Hot and warm confirmed exoplanets that belong to single-planet systems and orbit an only stellar component represent more than 40 \% of the observational sample\footnote{https://exoplanetarchive.ipac.caltech.edu/}. According to \citet{Zanardi2018}, the GR effects play a key role in the general dynamics of those systems, which makes them true laboratories of interest to study the behaviour of outer test particles with different orbital parameters. 

The general goal of the present research is to study the dynamical properties of outer test particles in the framework of the elliptical restricted three-body problem with GR effects. We are particularly interested in analyzing the role of the GR in the nodal librations of test particles located at the habitable zone (HZ) of the system, which evolve under the effects of an eccentric planetary-mass perturber with a semimajor axis of 0.1 au around a solar-mass star.   

The present work is organized as follows. In Sect. 2, we briefly present the analytical prescriptions used to carry out our investigation. In Sect. 3, we show a detailed analysis concerning nodal librations of HZ test particles in systems with different physical and orbital properties. In particular, we study the sensitivity of the results to the mass and the eccentricity of the inner perturber as well as to the eccentricity and the inclination of the HZ test particle. Moreover, we present results obtained from N-body experiments in order to test the robustness of the analytical theory. Finally, we describe the discussions and conclusions of our study in Sect. 4.

\section{Model - Analytical approach}

In this section, we present the model used to analyze the dynamical behavior of an outer test particle in the restricted elliptical three-body problem under the GR effects (RE3BP-GR). In particular, we describe the analytical approach derived by \citet{Zanardi2018}, who found an integral of motion associated with an outer test particle in the RE3BP-GR from the Hamiltonian up to the quadrupole level of the secular approximation obtained by \citet{Naoz2017} for an outer test particle in the restricted elliptical three-body problem (RE3BP). 

In fact, \citet{Ziglin1975} and \citet{Naoz2017} showed that the Hamiltonian of an outer test particle up to the quadrupole level of the secular approximation in the RE3BP is expressed by

\begin{eqnarray}
f_{\text{quad}} = \frac{\left(2 + 3e^2_{1}\right)\left(3\cos^2{i_{2}} - 1\right) + 15e^2_{1}\left(1 - \cos^2{i_{2}}\right)\cos{2\Omega_{2}}}{\left(1 - e^2_{2}\right)^{3/2}},
\label{eq:fquad}
\end{eqnarray}

\noindent{where} $e_1$ represents the inner perturber's eccentricity, and $e_2$, $i_2$, and $\Omega_2$ refer to the eccentricity, inclination, and ascending node longitude of the outer test particle, respectively. 

Later, \citet{Zanardi2018} showed that the RE3BP-GR for an outer test particle has associated an integral of motion $f$, which adopts the expression

\begin{eqnarray}
    f = f_{\text{quad}} + f_{\text{GR}},
    \label{eq:f}
\end{eqnarray}

\noindent{where} $f_{\text{quad}}$ is given by Eq.~\ref{eq:fquad} and $f_{\text{GR}}$ is expressed by

\begin{eqnarray}
    f_{\text{GR}} = \frac{48 k^2 \cos{i_{2}} \left(m_{1} + m_{\star}\right)^3 a_{2}^{7/2}\left(1 - e_{2}^2\right)^{1/2}}{m_{1}m_{\star}a_{1}^{9/2}c^2\left(1 - e_{1}^2\right)},
    \label{eq:fgr}
\end{eqnarray}

\noindent{where} $k^2$ is the gravitational constant, $c$ the speed of light, $m_{\star}$ and $m_1$ the mass of the star and the inner perturber, respectively, and $a_1$ and $a_2$ the semimajor axis of the inner perturber and the outer test particle, respectively.

Following \citet{Naoz2017}, if the outer particle’s ascending node longitude $\Omega_2$ is measured from the pericenter of the inner perturber, the precession of the inner perturber’s pericenter argument $\omega_1$ due to GR effects leads to a precession of $\Omega_2$. Thus, the temporal evolution of $\Omega_2$ in the RE3BP-GR is given by a combination between the secular evolution of $\Omega_2$ up to the quadrupole level of approximation and the precession of $\Omega_2$ induced by GR. According to the work carried out by \citet{Zanardi2018},

\begin{eqnarray}
    \frac{d{\Omega_{2}}}{dt} = \left(\frac{d{\Omega_{2}}}{dt}\right)_{\text{quad}} + \left(\frac{d{\Omega_{2}}}{dt}\right)_{\text{GR}},
    \label{eq:dodt}
\end{eqnarray}
\noindent{where} 

\begin{eqnarray}
    \left(\frac{d{\Omega_{2}}}{dt}\right)_{\text{quad}} = &-& \frac{m_{1}m{_\star}}{\left(m_{1}+m_{\star}\right)^2} n_{2}\left(\frac{a_{1}}{a_{2}}\right)^2 \\ \nonumber
    &\times&\frac{3\cos{i_{2}}\left(2 + 3e^2_{1} - 5e^2_{1}\cos{2{\Omega}_{2}} \right)}{8\left(1 - e^2_{2}\right)^2},
    \label{eq:omegaquad}
\end{eqnarray}



\noindent{being} $n_2 = k(m_1 + m_{\star})^{1/2}/a^{3/2}_2$, and

\begin{eqnarray}
\left(\frac{d{\Omega_{2}}}{dt}\right)_{\text{GR}} = -3k^3\frac{\left(m_{1} + m_{\star}\right)^{3/2}}{a_{1}^{5/2}c^2\left(1 - e^2_{1}\right)},
    \label{omegaGR}
\end{eqnarray}


\noindent{Now}, if we set ${\dot{\Omega}}_2 = 0$ in Eq.~\ref{eq:dodt}, the ascending node longitude's extreme values for libration trajectories of the outer test particle can be found. Thus, we obtain the corresponding value of $i_{2}$ that satisfies this condition as

\begin{eqnarray}
i_{2}^*=\arccos{\left(\frac{a_{2}^{7/2}\left(1 - e_{2}^2\right)^2A}{a_{1}^{9/2}\left(1 - e_{1}^2\right)\left(2 + 3e_{1}^2 - 5e_{1}^2\cos{2\Omega_{2}}\right)}\right)},
    \label{iestrella}
\end{eqnarray}

\noindent{where} $A$ is a constant given by

\begin{eqnarray}
    A = - \frac{8k^{2}\left( m_{1} + m_{\star} \right)^{3}}{c^{2}m_{1}m_{\star}}.
    \label{eq:A}
\end{eqnarray}

In this scenario of work, we can use the integral of motion $f$ given by Eq.~\ref{eq:f} to obtain the extreme values of the inclination $i_{2}$, which are reached when the ascending node longitude $\Omega_{2}$ adopts values of $\pm 90^{\circ}$. Following to \citet{Zanardi2018} and \citet{Zanardi2023}, the extreme inclinations $i_{2}^{\textrm{e}}$ that lead to nodal librations of the outer test particle are obtained from  

\begin{eqnarray}
    \alpha\cos^2{i_{2}^{\textrm{e}}} + \beta\cos{i_{2}^{\textrm{e}}} + \gamma = 0,
    \label{eq:cuadratica}
\end{eqnarray}

\noindent{where} $\alpha$ and $\beta$ are always given by

\begin{eqnarray}
    \alpha = 1 + 4e^{2}_{1},
    \label{alfa}
\end{eqnarray}
\begin{eqnarray}
    \beta = - \frac{A\left(1 - e^2_{2}\right)^{2}a^{7/2}_{2}}{\left(1 - e^2_{1}\right)a^{9/2}_{1}}.
    \label{beta}
\end{eqnarray}

\noindent{If} Eq.~\ref{iestrella} has solution at $\Omega = 0^{\circ}$, $\gamma$ is calculated by

\begin{eqnarray}
    \gamma = \frac{\beta^2}{4\left(1 - e^2_{1}\right)} - 5e^{2}_{1}.
    \label{gamma_viejo}
\end{eqnarray}

\noindent{On} the contrary, $\gamma$ is given by the following expression 

\begin{eqnarray}
    \gamma = \beta - \alpha,
    \label{gamma_nuevo}
\end{eqnarray}

\noindent{from} which, the maximum extreme inclination $i_{2,\textrm{max}}^{\textrm{e}}$ is always equal to 180$^{\circ}$ and the minimum extreme inclination adopts a simple form given by

\begin{eqnarray}
i_{2,\textrm{min}}^{\textrm{e}} = \textrm{arccos} \left(1 - \frac{\beta}{\alpha} \right).
\label{eq:i2min_nueva_cuadratica}
\end{eqnarray}

The resolution of Eq.~\ref{eq:cuadratica} allows us to derive the extreme values of the inclination that define the nodal libration region of an outer test particle in the RE3BP-GR. It is very important to remark that the coefficients $\alpha$, $\beta$ and $\gamma$ of that quadratic equation are functions of the orbital elements $a_{1}$, $e_{1}$, $a_{2}$, $e_{2}$ and $A$ parameter. According to this, the nodal libration region of an outer test particle in the RE3BP-GR strongly depends on the orbital and physical properties of the bodies that compose the system under study.

\section{Results}

In this section, we analyze the nodal librations of an outer test particle in the RE3BP-GR. In particular, we assume that all the systems of work are composed of a solar-mass star, an inner perturber with a semimajor axis $a_1 = 0.1$ au, and an outer test particle located at the HZ with a semimajor axis $a_2 = 1$ au. To carry out a detailed study about the evolution of these systems, our research is organized as follow. First, we use the analytical approach described in the previous section to analyze the nodal libration region of a HZ test particle that evolve under the effects of an inner Jupiter-mass planet for different values of $e_1$ and $e_2$. Then, the same analytical treatment is used in order to analyze the sensitivity of the nodal libration region to the mass of the inner perturber. Finally, we carry out a great set of N-body experiments with the aim of determining the robustness of our analytical results. 


\subsection{Sensitivity of the nodal libration region to the $e_1$ and $e_2$ values for a Jupiter-mass inner perturber}

\begin{figure}
   \centering
  \includegraphics[width=0.5
   \textwidth]{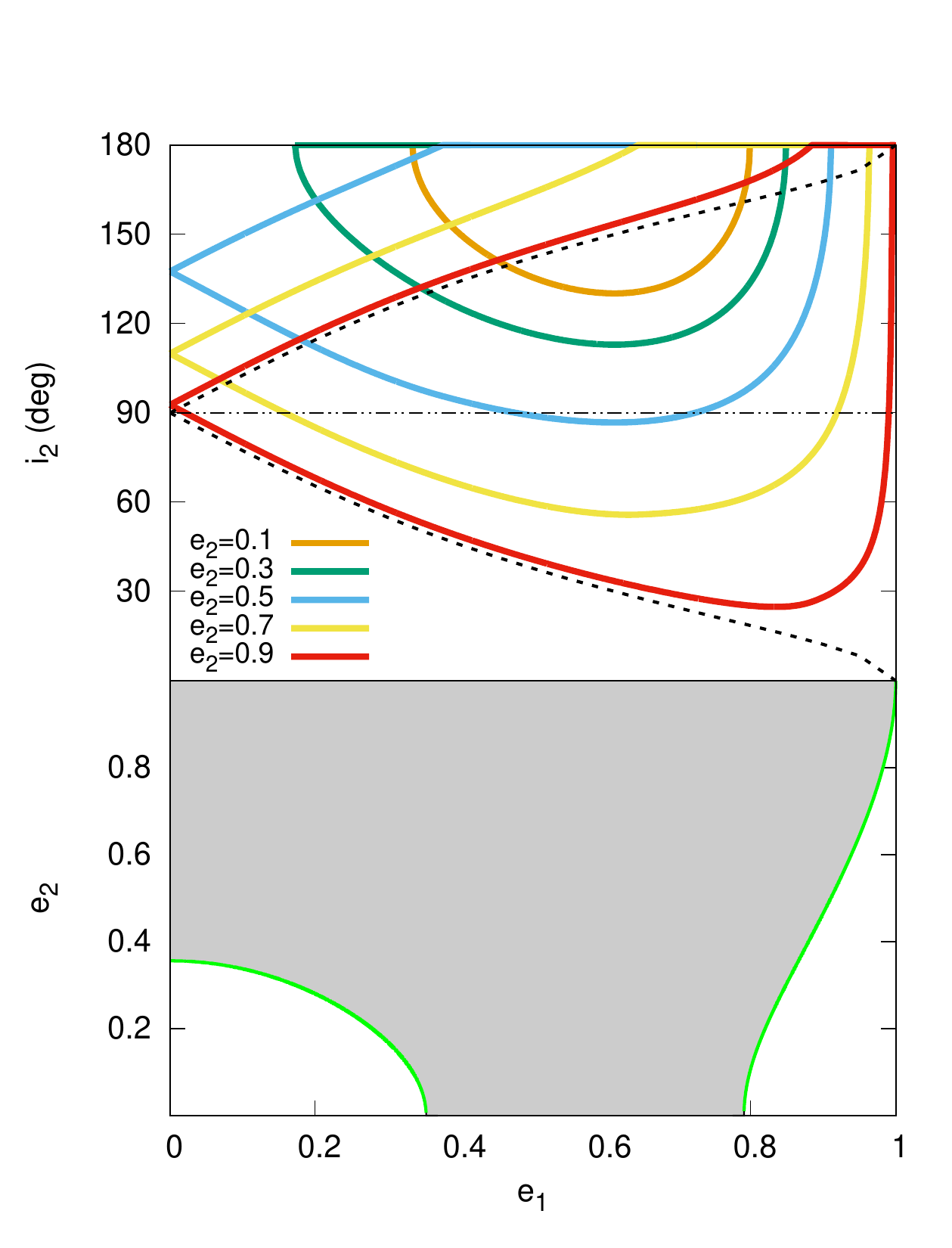}
   \caption{Top panel: Extreme values of $i_2$ that lead to nodal librations of the outer test particle with GR as a function of $e_1$ for different values of $e_2$. The black dotted curve represents such extreme inclinations in absence of GR. Bottom panel: values of $e_{2,\text{crit}}$ as a function of $e_1$ are illustrated by a green curve. The gray shaded region indicates the ($e_1$, $e_2$) pairs that can lead to nodal librations of the outer test particle for suitable values of $i_2$.
}
\label{fig:multiplot}
\end{figure}

By assuming a Jupiter-mass inner perturber, the top panel of Fig.~\ref{fig:multiplot} illustrates the extreme inclinations that produce nodal librations of the HZ test particle with GR as a function of $e_1$ for different values of $e_2$ as color curves. Moreover, the dotted black curve represents the extreme inclinations for nodal libration trajectories of the HZ test particle in absence of GR \citep{Ziglin1975}. From  this, several results of interest are evident. On the one hand, the range of prograde inclinations of the nodal libration region is reduced in comparison with that obtained without GR effects for any value of $e_1$ and $e_2$, which is consistent with that previously derived by \citet{Zanardi2018}. In fact, our results indicate that a HZ test particle with prograde inclinations can not experience nodal librations for $e_2 \lesssim$ 0.5 in this scenario of work. On the other hand, the greater the orbital eccentricity $e_2$, the wider the range of values associated with the inner planet's eccentricity $e_1$ that lead to nodal librations of the HZ test particle, which is in agreement with the results from \citet{Zanardi2018}. In particular, the top panel of Fig.~\ref{fig:multiplot} shows that, for a given $e_1$, the HZ test particle can evolve on nodal libration trajectories for values of $e_2$ greater than a critical value of the test particle's eccentricity ($e_{2,\textrm{crit}}$), for suitable values of $i_2$. If $e_1$ is fixed, the value of $e_{2,\textrm{crit}}$ is that for which the minimum and maximum extreme inclinations associated with nodal librations of the HZ test particle are both equal to 180$^{\circ}$. From Eq.~\ref{eq:i2min_nueva_cuadratica}, this condition requires that

\begin{eqnarray}
 -1 = 1 - \frac{\beta}{\alpha},
  \label{eq:e2vse1_1}
\end{eqnarray}

\noindent{which} leads to the solution

\begin{eqnarray}
    e_{2,\textrm{crit}}=\sqrt{1 - \left(\frac{4\left(1 - e^2_1\right)^2\left(1 + 4e^2_1 \right)^2 a^9_1}{A^2a^7_2}\right)^{1/4}}.
    \label{eq:e2crit_vs_e1}
\end{eqnarray}

\noindent{The} green curve in the bottom panel of Fig.~\ref{fig:multiplot} illustrates the values of $e_{2,\textrm{crit}}$ as a function of $e_1$ for our scenario of work. The gray shaded region above the curve represents the possible values of $e_2$ that lead to nodal librations of the HZ test particle for a given $e_1$ and suitable values of $i_2$. It is very interesting to note that an inner Jupiter-mass planet allows the HZ test particle to evolve on nodal libration trajectories for any value of $e_2$ and an appropriate combination of $e_1$ and $i_2$. 

\begin{figure*}
 \centering
\includegraphics[angle=360, width=1.0
   \textwidth]{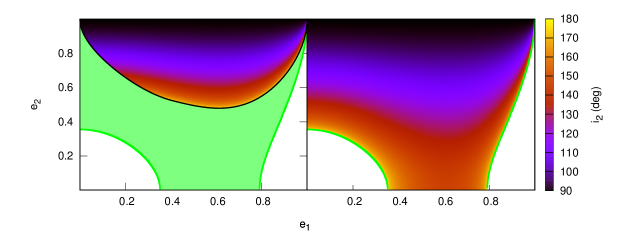}
\caption{
Left panel: Values of $e_{2,\text{crit}}$ and $e_{2,i^{\text{e}}_{\textrm{2,min}}=90^{\circ}}$ as a function of $e_1$ are illustrated by a green and a black curve, respectively. The green shaded region represents the pairs ($e_1$, $e_2$) that lead to nodal librations with purely retrograde orbits. The color code illustrates the corresponding value of $i^{\textrm{lim}}_2(\Omega_2=\pm 90^{\circ})$. Right panel: The color code represents $i_{2}( \Omega_{2}=\pm 90^{\circ}, \Delta i_2 = 0^{\circ})$ for a given pair ($e_1$, $e_2$). The green curve has the same reference as in the left panel.} 
\label{fig:multiplot_1x2_1jup_BIS}   
\end{figure*}

In agreement with \citet{Zanardi2018}, we find two different regimes of nodal librations for the HZ test particle, which depend on the evolution of $i_2$. On the one hand, nodal librations associated with purely retrograde orbits. On the other hand, nodal librations correlated with flips of the  orbital plane from prograde to retrograde and back again. From the top panel of Fig.~\ref{fig:multiplot}, it is possible to find a value of $e_2$ for each $e_1$ where the minimum extreme inclination of the nodal libration region is equal to 90$^{\circ}$. Such value is called as $e_{2,i^{\text{e}}_{\textrm{2,min}}=90^{\circ}}$. For a given $e_1$, the value of $e_{2,i^{\text{e}}_{\textrm{2,min}}=90^{\circ}}$ is that for which the minimum solution of the Eq.~\ref{eq:cuadratica} is equal to 0. If Eq.~\ref{iestrella} has solution at $\Omega_2 = 0^{\circ}$, the condition $i^{\text{e}}_{\textrm{2,min}}=90^{\circ}$ requires to solve the following equation

\begin{eqnarray}
    - \beta + (\beta^{2} - 4 \alpha \gamma)^{1/2} = 0,
    \label{eq:sol_min}
\end{eqnarray}

\noindent{with} $\gamma$ given by Eq.~\ref{gamma_viejo}, which allows us to obtain 

\begin{eqnarray}
    e_{2,i^{\text{e}}_{\textrm{2,min}}=90^{\circ}} = \sqrt{1 - \left(\frac{20  e^2_1 \left(1 - e^2_1 \right)^{3} a^9_1}{A^2a^7_2}\right)^{1/4}}.
    \label{eq:lim_flips}
\end{eqnarray}

\noindent{If} there is no solution for Eq.~\ref{iestrella} when $\Omega_2 = 0^{\circ}$, Eq.~\ref{eq:i2min_nueva_cuadratica} must be evaluated at $i^{\text{e}}_{\textrm{2,min}}=90^{\circ}$, which leads us to the equation

\begin{eqnarray}
    \beta - \alpha = 0,
    \label{eq:sol_min_nueva}
\end{eqnarray}

\noindent{obtaining}

\begin{eqnarray}
    e_{2,i^{\text{e}}_{\textrm{2,min}}=90^{\circ}} = \sqrt{1 - \left(\frac{\left(1-e^2_1\right)^2 \left(1 + 4 e^2_1 \right)^2 a^9_1}{A^2a^7_2}\right)^{1/4}}.
    \label{eq:lim_flips_nueva}
\end{eqnarray}

\noindent{Equating} Eqs.~\ref{eq:lim_flips} and \ref{eq:lim_flips_nueva}, it is possible to verify that both of them give the same $e_{2,i^{\text{e}}_{\textrm{2,min}}=90^{\circ}}$ for a value of $e_1$ = $\sqrt{1/6}$. Thus, the values of $e_{2,i^{\text{e}}_{\textrm{2,min}}=90^{\circ}}$ as a function of $e_1$ must be calculated using Eq.~\ref{eq:lim_flips} for $e_1 \leq$ $\sqrt{1/6}$, and Eq.~\ref{eq:lim_flips_nueva} for $e_1 >$ $\sqrt{1/6}$. It is important to mention two important points related to this discussion. On the one hand, Eqs.~\ref{eq:lim_flips} and \ref{eq:lim_flips_nueva} give the same $e_{2,i^{\text{e}}_{\textrm{2,min}}=90^{\circ}}$ at $e_1$ = $\sqrt{1/6}$ regardless the masses $m_\star$ and $m_1$ associated with the the central star and the inner perturber, respectively. On the other hand, the value of $e_{2,i^{\text{e}}_{\textrm{2,min}}=90^{\circ}}$ at $e_1$ = $\sqrt{1/6}$ does depend on $m_\star$ and $m_1$. These comments will be very important for our analysis of the Sect. 3.2, which will be associated with the sensitivity of the results to the inner perturber's mass.   

\noindent{The values of} $e_{2,i^{\text{e}}_{\textrm{2,min}}=90^{\circ}}$ as a function of $e_1$ are illustrated in the left panel of Fig.~\ref{fig:multiplot_1x2_1jup_BIS} by a black curve. Moreover, in both panels of such figure, the green curve represents the values of $e_{\textrm{2,crit}}$ previously derived from Eq.~\ref{eq:e2crit_vs_e1}.
For a given $e_1$, HZ test particles with orbital eccentricities $e_{2,\textrm{crit}} < e_2 < e_{2,i^{\text{e}}_{\textrm{2,min}}=90^{\circ}}$ experience nodal librations on purely retrograde orbits, since the minimum and maximum inclinations have retrograde values for any trajectory within the nodal libration region. These ($e_1$, $e_2$) pairs are illustrated by the green shaded region in the left panel of Fig.~\ref{fig:multiplot_1x2_1jup_BIS}. For $e_2 > e_{2,i^{\text{e}}_{\textrm{2,min}}=90^{\circ}}$, the minimum and maximum extreme inclinations of the nodal libration region have always prograde and retrograde values, respectively. In this case, a HZ test particle can experience nodal librations on purely retrograde orbits or orbital flips depending on the minimum and maximum inclinations of its evolutionary trajectory, which are associated with $\Omega_2 = \pm 90^{\circ}$. We call such values of the HZ test particle's orbital inclination as $i_2(\Omega_2=\pm 90^{\circ})$. For a prograde value of $i_2(\Omega_2=\pm 90^{\circ})$ within the nodal libration region when $e_2 > e_{2,i^{\text{e}}_{\textrm{2,min}}=90^{\circ}}$, a HZ test particle experiences librations of the ascending node longitude $\Omega_2$ together with orbital flips, since the extremes of $\Omega_2$ are always obtained for retrograde values of the inclination $i_2$ (Eq.~\ref{iestrella}). Thus, this class of HZ test particles shows oscillations of $\Omega_2$ correlated with flips of the orbital plane from prograde to retrograde and back again. For a retrograde value of $i_2(\Omega_2=\pm 90^{\circ})$ within the nodal libration region when $e_2 > e_{2,i^{\text{e}}_{\textrm{2,min}}=90^{\circ}}$, the specification of the nodal libration regime is more complex, since it is necessary to determine if the other value of $i_2(\Omega_2=\pm 90^{\circ})$ over the evolutionary trajectory is prograde or retrograde. To do this, we make use of the integral of motion $f$ given by Eq.~\ref{eq:f}, which is conserved over the evolutionary trajectory of each HZ test particle. For a retrograde value of $i_2(\Omega_2=\pm 90^{\circ})$ between 90$^{\circ}$ and the maximum extreme inclination of the nodal libration region, this procedure allows us to calculate the other value of $i_2(\Omega_2=\pm 90^{\circ})$ associated with the evolutionary trajectory of a HZ test particle for each pair ($e_1$, $e_2$) above the black curve of the left panel of Fig.~\ref{fig:multiplot_1x2_1jup_BIS}. By assuming that $i^{\dag}_2$ is the known value of $i_2(\Omega_2=\pm 90^{\circ})$, the other $i_2(\Omega_2=\pm 90^{\circ})$ is obtained by solving the following quadratic equation

\begin{eqnarray}
    \alpha\cos^2{i_2(\Omega_2=\pm 90^{\circ})} + \beta\cos{i_2(\Omega_2=\pm 90^{\circ})} + \gamma^{\dag} = 0,
    \label{eq:cuadratica_tray_indv}
\end{eqnarray}

\noindent{where} $\alpha$ and $\beta$ are always given by Eqs.~\ref{alfa} and \ref{beta}, respectively, and $\gamma^{\dag}$ adopts the expression

\begin{eqnarray}
    \gamma^{\dag} = - \alpha\cos^2{i^{\dag}_2} - \beta\cos{i^{\dag}_2}.
    \label{eq:gamma_daga}
\end{eqnarray}

\noindent{From} this, we find that there is a limit retrograde value of $i_2(\Omega_2=\pm 90^{\circ})$ for $e_2 > e_{2,i^{\text{e}}_{\textrm{2,min}}=90^{\circ}}$ called $i^{\textrm{lim}}_2(\Omega_2=\pm 90^{\circ})$, which divides the two nodal libration regimes. In fact, if $90^{\circ} < i_2(\Omega_2=\pm 90^{\circ}) < i^{\textrm{lim}}_2(\Omega_2=\pm 90^{\circ})$, the trajectory of nodal libration is associated with purely retrograde orbits, while if $i^{\textrm{lim}}_2(\Omega_2=\pm 90^{\circ}) < i_2(\Omega_2=\pm 90^{\circ}) \leq i^{\textrm{e}}_{2,\textrm{max}}$, the nodal libration is correlated with flips of the orbital plane between prograde and retrograde and back again. The color code in the left panel of Fig.~\ref{fig:multiplot_1x2_1jup_BIS} illustrates the value of $i^{\textrm{lim}}_2(\Omega_2=\pm 90^{\circ})$ for each pair $(e_1, e_2)$ above the black curve referred to $e_{2,i^{\text{e}}_{\textrm{2,min}}=90^{\circ}}$.   

According to this analysis, it is worth mentioning that a given inner perturber only allows a HZ test particle to experience nodal librations with orbital flips for $e_2$ greater than the minimum value of the curve associated with  $e_{2,i^{\text{e}}_{\textrm{2,min}}=90^{\circ}}$ in the ($e_1$, $e_2$) plane, which is constructed from Eqs.~\ref{eq:lim_flips} and \ref{eq:lim_flips_nueva} for $e_1$ less than and greater than $\sqrt{1/6}$, respectively. To determine such a minimum value, it is necessary to analyze Eqs.~\ref{eq:lim_flips} and \ref{eq:lim_flips_nueva} individually. On the one hand, the derivative of Eq.~\ref{eq:lim_flips} respect to $e_1$ is given by



\begin{eqnarray}
\frac{\textrm{d}e_{2,i^{\text{e}}_{\textrm{2,min}}=90^{\circ}}}{\textrm{d}e_1}=-\frac{1}{4}\left(\frac{20 a^{9}_{1}}{ A^{2} a^{7}_2}\right)^{1/4} \frac{(1 - 4 e^2_1)}{\sqrt{e_1(1-e^{2}_1)^{1/2}} e_{2,i^{\text{e}}_{\textrm{2,min}}=90^{\circ}}} ,
    \label{eq:deri_e2imin90_1}
\end{eqnarray}

\noindent{where} $e_{2,i^{\text{e}}_{\textrm{2,min}}=90^{\circ}}$ refers to  Eq.~\ref{eq:lim_flips}. From this, Eq.~\ref{eq:lim_flips} 
has a minimum value at $e_1$ = 0.5, which is outside the range of validity of such an equation. On the other hand, the derivative of Eq.~\ref{eq:lim_flips_nueva} adopts the following expression

\begin{eqnarray}
\frac{\textrm{d}e_{2,i^{\text{e}}_{\textrm{2,min}}=90^{\circ}}}{\textrm{d}e_1}= -\frac{1}{2}\left(\frac{a^{9}_{1}}{A^2a^{7}_2}\right)^{1/4} \frac{e_1  (3 - 8e^2_1)}{\sqrt{1+3e^{2}_1-4e_{1}^{4}}e_{2,i^{\text{e}}_{\textrm{2,min}}=90^{\circ}}}, 
    \label{eq:deri_e2imin90_2}
\end{eqnarray}


\noindent{where} $e_{2,i^{\text{e}}_{\textrm{2,min}}=90^{\circ}}$ refers to  Eq.~\ref{eq:lim_flips_nueva}. According to this, the minimum of Eq.~\ref{eq:lim_flips_nueva} is obtained when $e_1$ = $\sqrt{3/8}$, which is within its range of validity. Thus, the minimum value of the curve associated with  $e_{2,i^{\text{e}}_{\textrm{2,min}}=90^{\circ}}$ in the ($e_1$, $e_2$) plane must always be calculated by evaluating Eq.~\ref{eq:lim_flips_nueva} at $e_1$ = $\sqrt{3/8}$, which is valid for any value of the masses $m_\star$ and $m_1$ associated with the central star and the inner perturber, respectively. For the particular case of an inner Jupiter-mass planet around a solar-mass star, the minimum value of $e_{2,i^{\text{e}}_{\textrm{2,min}}=90^{\circ}}$ is equal to 0.458. According to this, HZ test particles with $e_2 <$ 0.458 can not experience nodal librations correlated with orbital flips for any set of parameters ($e_1$, $i_2$). This detailed analysis is consistent with that result initially illustrated in Fig.~\ref{fig:multiplot}, which indicated that a HZ test particle with prograde inclinations can not experience nodal librations for $e_2 \lesssim$ 0.5 in
the present scenario of work.  

\begin{figure*}
\centering
\includegraphics[angle=360, width=1.0
   \textwidth]{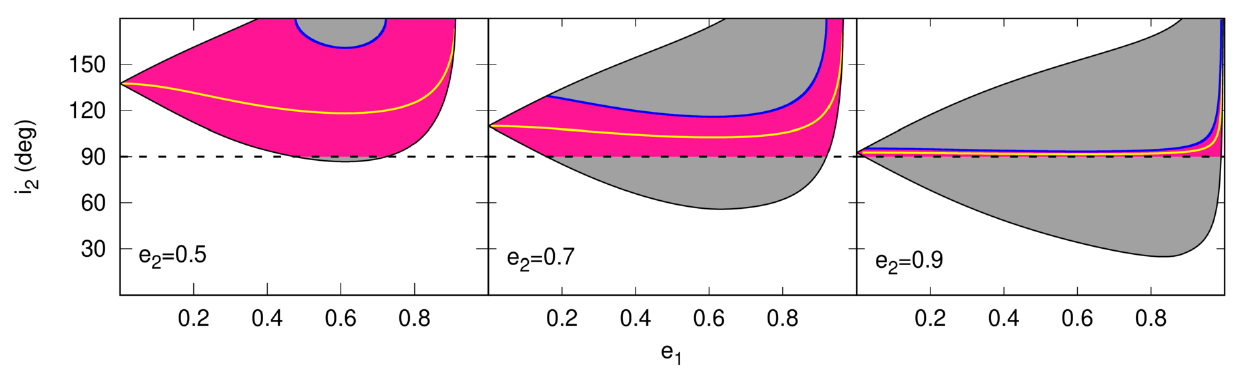} 
\caption{
Nodal libration region of an outer particle in the ($e_1$,$i_2$) plane for values of $e_2$ = 0.5 (left), 0.7 (middle), and 0.9 (right). The gray and dark pink shaded regions represent the ($e_1$,$i_2$) pairs that lead to nodal librations with orbital flips and purely retrograde orbits, respectively. The blue and yellow curves illustrate $i^{\textrm{lim}}_2(\Omega_2=\pm 90^{\circ})$ and $i_{2}(\Omega_{2}=\pm 90^{\circ}, \Delta i_2 = 0^{\circ})$, respectively.
}
\label{fig:fc_1Mj_multiplot}
\end{figure*}

The calculation of $i^{\textrm{lim}}_2(\Omega_2=\pm 90^{\circ})$ in the left panel of Fig.~\ref{fig:multiplot_1x2_1jup_BIS} gives important information since it allows us to visualize the two nodal libration regimes in the ($e_1$, $i_2$) plane for a given value of $e_2$. From this, Fig.~\ref{fig:fc_1Mj_multiplot} illustrates $i^{\textrm{lim}}_2(\Omega_2=\pm 90^{\circ})$ as a function of $e_1$ by blue curves for values of $e_2$ of 0.5 (left panel), 0.7 (middle panel), and 0.9 (right panel). In each panel, the gray shaded region indicates the pairs $(e_1, i_2(\Omega_2=\pm 90^{\circ}))$ that lead to nodal librations with orbital flips, while the dark pink shaded region represents the pairs $(e_1, i_2(\Omega_2=\pm 90^{\circ}))$ that lead to nodal librations on purely retrograde orbits. According to this, it is evident that the greater the HZ test particle's eccentricity $e_2$, the greater the range of $i_2(\Omega_2=\pm 90^{\circ})$ that produce nodal libration trajectories correlated with orbital flips for a given $e_1$ in our scenario of study. 

Finally, it is worth remarking that the above analysis allowed us to find peculiar purely retrograde trajectories within the nodal libration region for which the minimum and maximum values of the HZ test particle's inclination are equal. We call such values as $i_{2}(\Omega_{2}=\pm 90^{\circ}, \Delta i_{2} = 0^{\circ})$. Given the correlation between the inclination and the ascending node longitude, a HZ test particle with an orbital inclination $i_{2}(\Omega_{2}=\pm 90^{\circ}, \Delta i_2 = 0^{\circ})$ evolves in time with constant values of $i_2$ and $\Omega_2$. The right panel of Fig.~\ref{fig:multiplot_1x2_1jup_BIS} illustrates the value of $i_{2}(\Omega_{2}=\pm 90^{\circ}, \Delta i_2 = 0^{\circ})$ for each pair $(e_1, e_2)$ as a color code. Moreover, the values of $i_{2}(\Omega_{2}=\pm 90^{\circ}, \Delta i_2 = 0^{\circ})$ are represented by a yellow curve as a function of $e_1$ for each $e_2$ considered in the panels of Fig~\ref{fig:fc_1Mj_multiplot}. In general terms, the greater the HZ test particle's eccentricity $e_2$, the smaller the value of $i_{2}( \Omega_{2}=\pm 90^{\circ}, \Delta i_2 = 0^{\circ})$. 

From Figs.~\ref{fig:multiplot_1x2_1jup_BIS} and \ref{fig:fc_1Mj_multiplot}, it is very interesting to develop a discussion about the dynamical evolution of HZ test particles with extremely eccentric orbits. In fact, for very high values of $e_2$, the test particle can only experience nodal librations on purely retrograde orbits for values of $i_2(\Omega_2=\pm 90^{\circ})$ close to 90$^{\circ}$. Moreover, the libration amplitude associated with the inclination of those quasi-polar orbits is close to (or even equal to) zero, according to the yellow curve illustrated in the right panel of Fig.~\ref{fig:fc_1Mj_multiplot}. These results are no longer valid for an extremely eccentric inner Jupiter-mass perturber, since the range of $i_2(\Omega_2=\pm 90^{\circ})$ that lead to nodal librations with purely retrograde orbits and the values of $i_{2}(\Omega_{2}=\pm 90^{\circ}, \Delta i_2 = 0^{\circ})$ increase with $e_1$. We want to remark that these conclusions should be carefully interpreted since the present analysis is based on a secular theory up to the quadrupole level, which is not appropriate to describe the dynamical behavior of HZ test particles with extremely eccentric orbits. Indeed, for very high values of $e_2$, non-secular and  higher order secular terms of the disturbing function should play an important role in the dynamical evolution of the particles under consideration. Beyond this, we decide to make use of our analytical approximation in order to derive dynamical criteria that lead to nodal librations of a HZ test particle for the full range of eccentricities $e_2$ for completeness reasons.

\subsection{Sensitivity of the nodal libration region to the inner perturber's mass}

Here, we analyze the nodal librations of a HZ test particle that evolves under the effects of an inner planet of mass $m_1$ around a solar-mass star in the RE3BP-GR. In particular, we study the sensitivity of the results to $m_1$, adopting values ranging from terrestrial-like planets to super-Jupiters. 

\begin{figure}
\centering
\includegraphics[angle=360, width=0.5
   \textwidth]{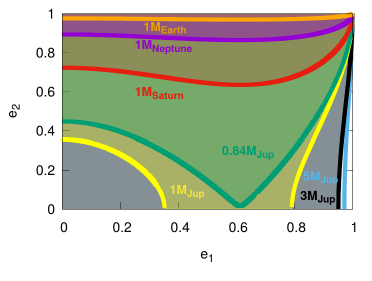} 
\caption[aver]{Values of $e_{2,\text{crit}}$ as a function of $e_1$ are illustrated by color curves for different $m_1$. The corresponding color shaded regions represent the values of $e_1$ and $e_2$ that lead to nodal librations for a suitable $i_2$. 
}
\label{fig:e2vse1_opc2_fc}
\end{figure}

Figure~\ref{fig:e2vse1_opc2_fc} illustrates the values of $e_{2,\textrm{crit}}$ as a function of $e_1$ by color curves, which were derived from Eq.~\ref{eq:e2crit_vs_e1} for different masses of the inner perturber. The shaded region above each curve associated with a given $m_1$ represents the ($e_1$, $e_2$) pairs that lead to nodal librations of the HZ test particle for suitable values of $i_2$. On the one hand, our results show that nodal libration trajectories of the HZ test particle are possible for any value of the mass $m_1$ and the eccentricity $e_1$ of the inner planet for suitable values of the eccentricity $e_2$ and inclination $i_2$. On the other hand, we find that the more massive the inner planet, the greater the nodal libration region in the ($e_1$, $e_2$) plane. 

\begin{figure*}
\centering
\includegraphics[angle=360, width=1
   \textwidth] {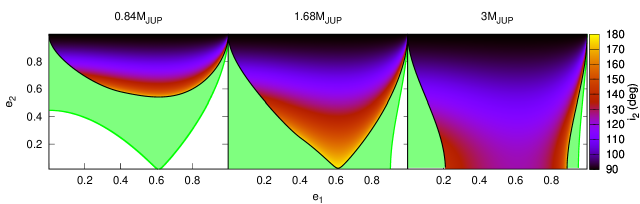} 
\caption[aver]{Sets of ($e_1$, $e_2$) values associated with the nodal libration region of an outer test particle for a perturber of 0.84 M$_{\text{Jup}}$ (left), 1.68 M$_{\text{Jup}}$ (middle) and 3 M$_{\text{Jup}}$ (right). The curves, regions and the color code shown in this figure are described in the caption associated with the left panel of Fig.~\ref{fig:multiplot_1x2_1jup_BIS}. 
}
\label{fig:multiplot_e2vse2_regioneslibr_variasmasas}
\end{figure*}

\begin{figure*}
\centering
\includegraphics[angle=360, width=1
   \textwidth]{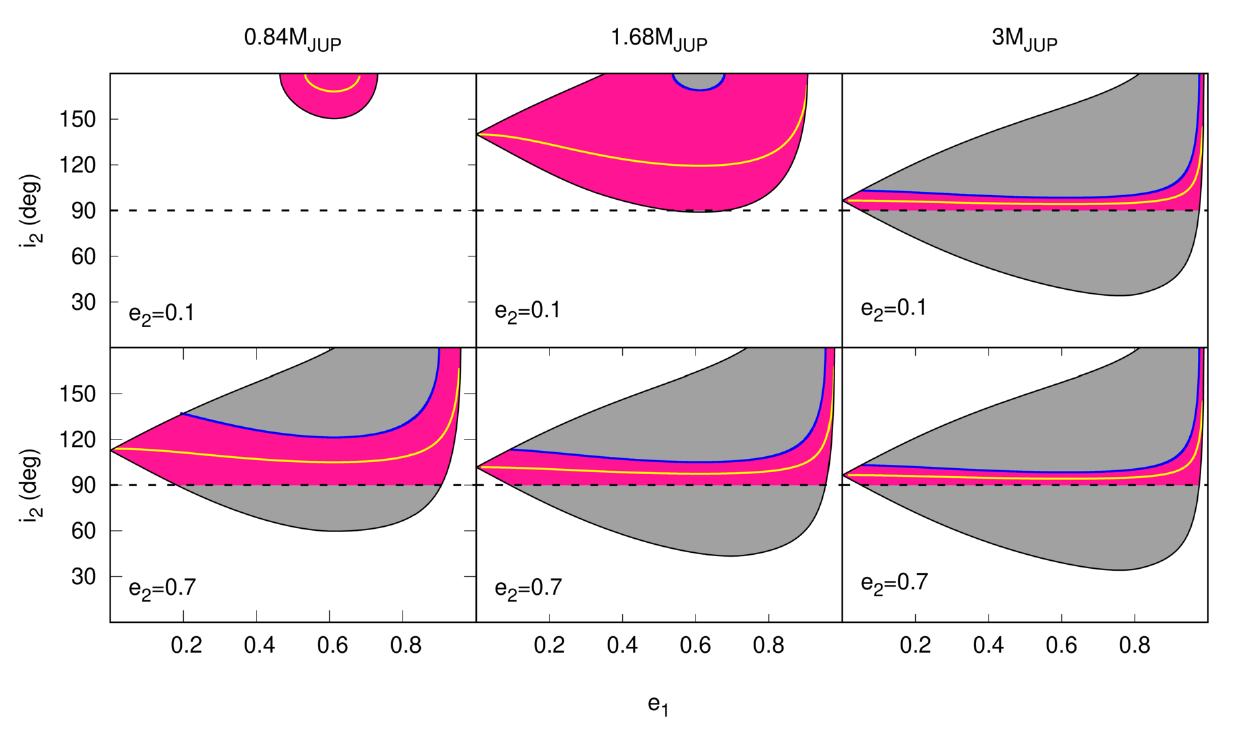} 
\caption[aver]{
Nodal libration region of an outer test particle in a ($e_1$, $i_2$) plane for $e_2$ = 0.1 (top) and 0.7 (bottom), and $m_1$ = 0.84 M$_{\text{Jup}}$ (left), 1.68 M$_{\text{Jup}}$ (middle) and 3 M$_{\text{Jup}}$ (right). The curves and regions shown in this figure are described in the caption associated with Fig.~\ref{fig:fc_1Mj_multiplot}. 
}
\label{fig:multiplot_2x3_1sat_3jup}
\end{figure*}

From terrestrial-like planets to sub-Jupiters, Fig.~\ref{fig:e2vse1_opc2_fc} allows us to observe that there is a minimum value of $e_2$ below which nodal librations of the HZ test particle are not possible for any $e_1$. 
To find such a minimum value of $e_{2}$ for each $m_1$, we derive Eq.~\ref{eq:e2crit_vs_e1} respect to $e_1$, obtaining 

\begin{eqnarray}
    \frac{\textrm{d}e_{2,\textrm{crit}}}{\textrm{d}e_1}=-\frac{1}{2}\left(\frac{4a^{9}_{1}}{ A^{2} a^{7}_2}\right)^{1/4} \frac{ e_1 (3 - 8 e^2_1)}{ \sqrt{1+3e^{2}_1-4e_{1}^{4}} e_{2,\textrm{crit}}},
    \label{eq:deri_e2vse1}
\end{eqnarray}


\noindent{where} $e_{2,\textrm{crit}}$ in the denominator corresponds to Eq.~\ref{eq:e2crit_vs_e1}. It is worth noting that this derivative vanishes for a value of $e_1 =$ $\sqrt{3/8}$  for any $m_1$, since the dependence on this physical variable only is given from the $A$ parameter defined in Eq.~\ref{eq:A}. According to this, an inner Earth-, Neptune-, Saturn-mass planet only allows a HZ test particle to experience nodal librations for $e_2$ greater than 0.969, 0.864, and 0.634 respectively. From these scenarios of work, the more massive the inner planet, the greater the range of values of the HZ test particle's eccentricity that produce nodal librations. This correlation between $m_1$ and $e_2$ shows that there is a limit mass $m_{1,\textrm{lim}}$ above which it is possible to find nodal libration trajectories for any value of the HZ test particle's eccentricity $e_2$, and suitable values of $e_1$ and $i_2$. To calculate $m_{1,\textrm{lim}}$, it is necessary that $e_{2,\textrm{crit}}$ = 0 in the minimum of Eq.~\ref{eq:e2crit_vs_e1} given by $e_1$ = $\sqrt{3/8}$. From this, a value of $m_{1,\textrm{lim}}$ = 0.84 M$_{\textrm{Jup}}$ is derived. The values of $e_{2,\textrm{crit}}$ as a function of $e_1$ for the particular case of $m_{1,\textrm{lim}}$ are illustrated in Fig.~\ref{fig:e2vse1_opc2_fc} by a green curve. From these analyses,  inner Jupiter-like planets and super-Jupiters can produce nodal librations of a HZ test particle for any value of $e_2$ and a appropriate combination of $e_1$ and $i_2$.  

An important result of our research indicates that it is always possible to find a set of parameters ($e_1$, $e_2$, $i_2$) that lead to nodal librations of the HZ test particle both on purely retrograde orbits and with orbital flips for any value of the inner perturber's mass $m_1$.
It is worth remarking that this result is valid from sub-Earth-mass planets to super-Jupiters. 

From that discussed in Sect. 3.1, for a given $m_1$, Eq.~\ref{eq:lim_flips_nueva} evaluated at $e_1$ = $\sqrt{3/8}$ gives the minimum value of $e_2$ above which nodal librations with orbital flips of the HZ test particle are possible for an appropriate combination of $e_1$ and $i_2$. Following this procedure, an inner Earth-, Neptune-, Saturn-mass planet only allows a HZ test particle to experience nodal librations with orbital flips for $e_2$ greater than 0.978, 0.906, and 0.760, respectively. According to this, the more massive the inner perturber, the greater the range of values of $e_2$ that leads to nodal librations with orbital flips of the HZ test particle. Following this analysis, it is evident that there must exist a mass limit $m^{\S}_{1,\textrm{lim}}$ of the inner perturber above which nodal librations with orbital flips are possible for any value of $e_2$ and an appropriate combination of $e_1$ and $i_2$. The mass limit $m^{\S}_{1,\textrm{lim}}$ is that for which $e_{2,i^{\text{e}}_{\textrm{2,min}}=90^{\circ}} = 0$ in the minimum of Eq.~\ref{eq:lim_flips_nueva}. From this, we compute a value of $m^{\S}_{1,\textrm{lim}}$ equal to 1.68 M$_{\textrm{Jup}}$. It is interesting to note that an inner super-Jupiter allows a HZ test particle experience nodal librations with orbital flips for any $e_2$ and suitable values of $e_1$ and $i_2$.

In this line of analysis, we study the sensitivity of the nodal libration region associated with orbital flips in the ($e_1$, $i_2$) plane to the inner perturber's mass for a given value of $e_2$. To do this, we compare the nodal libration region of a HZ test particle produced by three different inner giant planets more massive than 0.84 M$_{\textrm{Jup}}$. We select the mass of the inner perturber in such a range since the HZ test particle can experience nodal librations for any value of $e_2$. From this, it is possible to analyze the dependence of the nodal libration region correlated with orbital flips on the inner perturber's mass both for low and high eccentricities of the HZ test particle.   

Figure~\ref{fig:multiplot_e2vse2_regioneslibr_variasmasas} illustrates the values of $e_{2,\textrm{crit}}$ (black curve) and $e_{2,i^{\text{e}}_{\textrm{2,min}}=90^{\circ}}$ (green curve) as a function of $e_1$ for an inner perturber of 0.84 M$_{\textrm{Jup}}$ (left panel), 1.68 M$_{\textrm{Jup}}$ (middle panel), and 3 M$_{\textrm{Jup}}$ (right panel). Like Fig.~\ref{fig:multiplot_1x2_1jup_BIS}, the green shaded region of each panel represents the pairs ($e_1$, $e_2$) that can produce nodal librations of the HZ test particle on purely retrograde orbits, while the color code above the green curve illustrates the limit retrograde value $i^{\textrm{lim}}_2(\Omega_{2} = \pm 90^{\circ})$ for each pair ($e_1$, $e_2$), which divides trajectories associated with purely retrograde orbits ($90^{\circ} < i_2(\Omega_2=\pm 90^{\circ}) < i^{\textrm{lim}}_2(\Omega_2= \pm 90^{\circ})$) and orbital flips ($i^{\textrm{lim}}_2(\Omega_2=\pm 90^{\circ}) < i_2(\Omega_2= \pm 90^{\circ}) \leq i^{\textrm{e}}_{2,\textrm{max}}$) within the nodal libration region. From these results, we construct Fig.~\ref{fig:multiplot_2x3_1sat_3jup}, which illustrates the nodal libration region in the ($e_1$, $i_2$) plane associated with purely retrograde orbits (dark pink) and orbital flips (gray) for an inner perturber of 0.84 M$_{\textrm{Jup}}$ (left panels), 1.68 M$_{\textrm{Jup}}$ (middle panels), and 3 M$_{\textrm{Jup}}$ (right panels) and values of $e_2$ of 0.1 (top panels) and 0.7 (bottom panels). The limit between both regimes of nodal libration is given by $i^{\textrm{lim}}_2(\Omega_2=\pm 90^{\circ})$, which is represented in each panel by a blue curve. From this, it is evident that the more massive the inner perturber, the greater the nodal libration region associated with orbital flips in the ($e_1$, $i_2$) plane for a given value of $e_2$.

Finally, the yellow curve in each panel of Fig.\ref{fig:multiplot_2x3_1sat_3jup} represents the value of $i_{2}(\Omega_{2} = \pm 90^{\circ}, \Delta i_{2} = 0^{\circ})$ as a function of $e_1$, for which the libration amplitude of the HZ test particle's inclination is null throughout its evolution. According to this, for a given $e_2$, the more massive the inner perturber, the smaller the value of $i_{2}( \Omega_{2} = \pm 90^{\circ}, \Delta i_2 = 0^{\circ})$. This result indicates that an inner massive super-Jupiter allows HZ test particles to evolve on quasi-polar orbits with null libration amplitude of the inclination $i_2$ for low, moderate, and high values of the eccentricity $e_2$. This result is very interesting since it has important implications in the dynamical evolution and stability of polar planets in the HZ around solar-type stars.

As mentioned in Sect. 3.1,   the results derived for high $e_2$ values should be carefully interpreted due to the limitations of our analytical model based on a secular and quadrupolar Hamiltonian.

%

\subsection{Comparison with numerical experiments}

\begin{figure*}
\centering
\includegraphics[angle=360, width=1
   \textwidth]{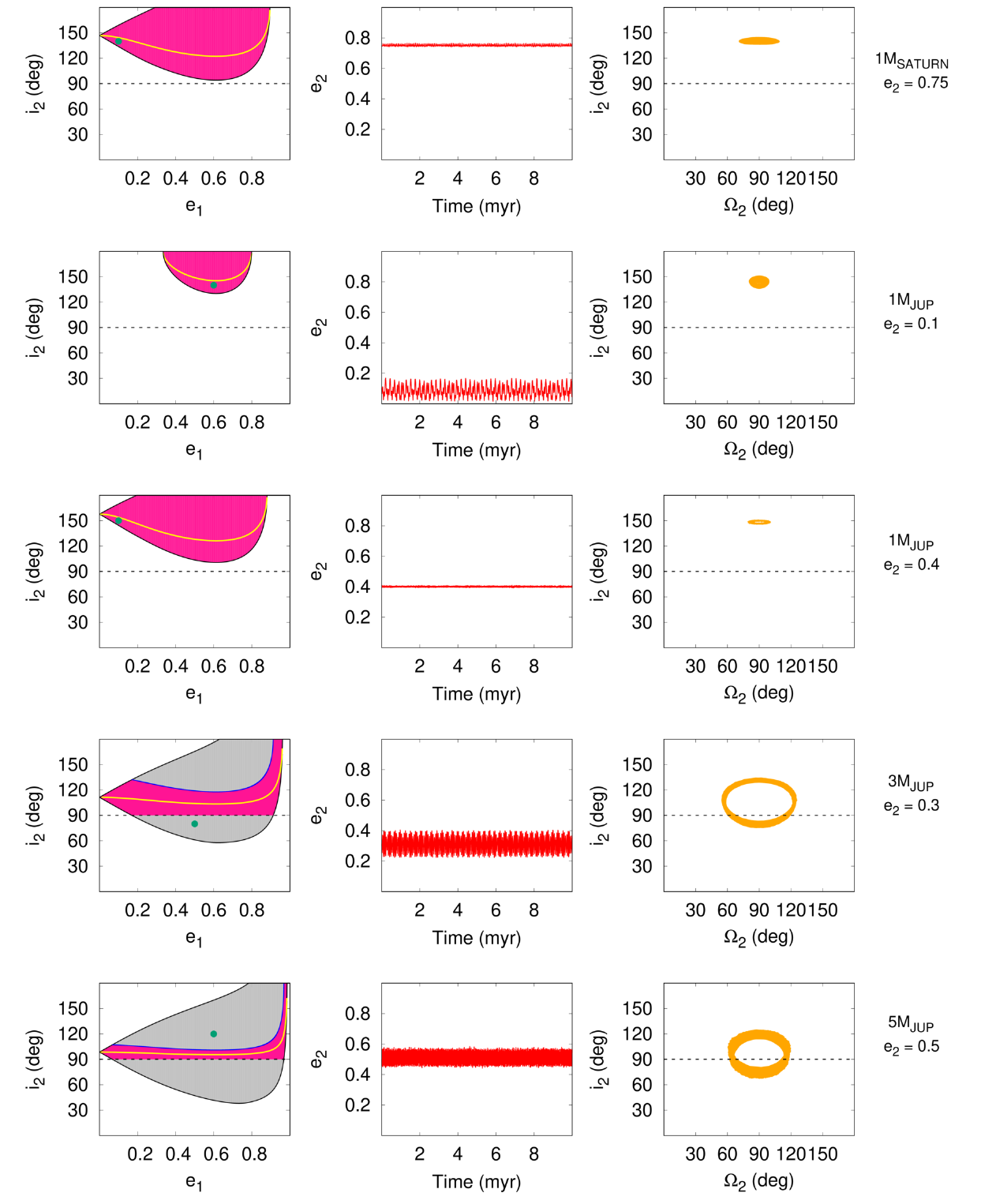} 
\caption[aver]{Nodal libration region in a ($e_1$, $i_2$) plane (left), temporal evolution of $e_2$ (middle), and evolutionary trajectory in a ($\Omega_2$, $i_2$) plane (right) of test particles resulting from N-body experiments with GR in different scenarios of work. The green circle in the left panels illustrates the value of $e_1$ and the initial $i_2$ at $\Omega_2 = 90^{\circ}$ of the simulated particles. The gray and dark pink shaded regions show the nodal libration region associated with orbital flips and purely retrograde orbits, respectively. Moreover, the blue and yellow curves represent $i^{\textrm{lim}}_2(\Omega_2=\pm 90^{\circ})$ and $i_{2}(\Omega_{2}=\pm 90^{\circ}, \Delta i_2 = 0^{\circ})$ as a function of $e_1$, respectively. The rows are numbered from 1 to 5 from top to bottom.
}
\label{fig:multiplotprueba}
\end{figure*}

Once the analytical criteria that lead to the generation of nodal librations of the HZ test particle have been defined, we test their robustness using numerical simulations. To do this, we constructed a modified version of the well-known MERCURY code \citep{Chambers1999}, by including GR effects from the correction proposed by \citet{Anderson1975}, which is given by

\begin{eqnarray}
\Delta{{\ddot{\bf r}}} = \frac{k^{2}m_{\star}}{c^2 r^3} \Bigg \{ \left( \frac{4k^{2}m_{\star}}{r} - {\bf v}\cdot {\bf v} \right) {\bf r} + 4\left({\bf r}\cdot {\bf v} \right){\bf v} \Bigg \},
\label{eq:GR-Anderson}
\end{eqnarray}
where \textbf{r} and \textbf{v} are the astrocentric position and velocity vectors, respectively, and $r$ = |\textbf{r}|. It is important to highlight that we are working with a first order post-Newtonian approximation. Finally, we carried out all numerical experiments making use of the Bulirsch–Stoer algorithm adopting an accuracy parameter of 10$^{-12}$. In order to carry out a correct comparison between the analytical and numerical results, we remark that the orbital elements of the outer test particle always must be referenced to the barycenter and invariant plane of the system, where x-axis coincides with the pericenter of the inner perturber. Since GR effects leads to a precession of the inner perturber’s argument of pericenter $\omega_1$, the ascending node longitude of the test particle $\Omega_2$ is measured respect to a rotating system.

We were not able to find examples of HZ test particles with nodal librations that survive a 10 Myr full N-body experiment for an inner perturber less massive than one Neptune mass.
In fact, according to the criteria discussed in Sect. 3.2 from a secular theory to quadrupole level, such an inner perturber can only produce nodal librations of a HZ test particle with an eccentricity $e_2$ greater than about 0.86. For such high values of $e_2$, our numerical experiments show that the dynamical evolution of the HZ test particle is governed by close encounters with the inner perturber, for which its typical final fate is a collision with the planet or the central star, or an ejection from the system after a few hundred thousand years.

For an inner Saturn-mass planet, the results of the secular theory shown in Sect. 3.2 indicate that nodal librations of the HZ test particle require values of $e_2$ greater than about 0.63. According to this, we carried out N-body simulations for values of $e_2$ of 0.65, 0.7, 0.75, 0.8, 0.85 and 0.9, and values of $e_1$ ranging between 0.1 and 0.8. Our study shows that nodal librations of the HZ test particle only survive an N-body treatment for $e_1$ = 0.1 and $e_2$ = 0.75. In this case, the eccentricity and inclination of the HZ test particle evolve keeping a value close to the initial one of 0.75 and 140$^{\circ}$, respectively, while the ascending node longitude $\Omega_2$ librates throughout the total integration time of 10 Myr, as observed in the row 1 of Fig.~\ref{fig:multiplotprueba}. This behavior is in agreement with that derived from the secular theory described in the present research, which allows us to see that the nodal librations are only correlated with purely retrograde orbits of the HZ test particle for $m_1$ = 1 M$_{\textrm{Sat}}$, $e_1$ = 0.1, $e_2$ = 0.75. For greater values of $e_1$ and $e_2$, our numerical experiments show that the HZ test particle’s eccentricity experiences a chaotic evolution, which leads the particle to collide with the planet or the central star, or to be ejected from the system after a few million years.  

\begin{figure*}
\centering
\includegraphics[angle=360, width=1
   \textwidth]{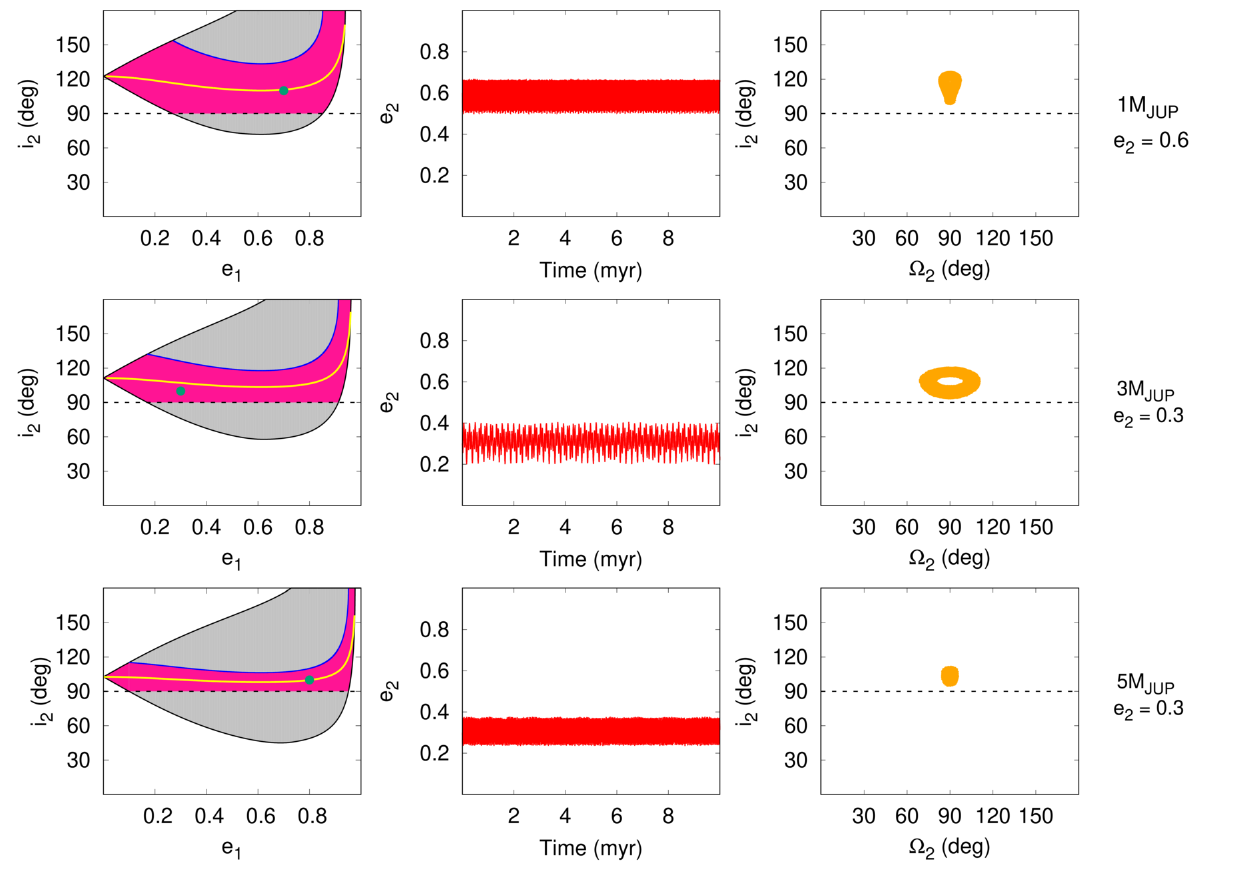} 
\caption[aver]{Nodal libration region in a ($e_1$, $i_2$) plane (left), temporal evolution of $e_2$ (middle), and evolutionary trajectory in a ($\Omega_2$, $i_2$) plane (right) of test particles with purely retrograde orbits resulting from N-body experiments with GR in different scenarios of work. The circles, curves and regions shown in this figure are described in the caption associated with Fig.~\ref{fig:multiplotprueba}.
}
\label{fig:fig8_multiplot_3x3_PR}
\end{figure*}

For an inner perturber more massive than 0.84 M$_{\textrm{Jup}}$, nodal librations of the HZ test particle are possible for any value of $e_2$, according to that derived in Sect. 3.2 on the basis on the secular approximation. According to this, we carried out N-body simulations for an inner perturber of 1 M$_{\textrm{Jup}}$, 3 M$_{\textrm{Jup}}$, and 5 M$_{\textrm{Jup}}$, by assuming a wide range of values associated with $e_2$ and $e_1$.

For an inner Jupiter-mass planet, we developed N-body simulations for values of $e_2$ of 0.1, 0.3, 0.4, 0.5 and 0.6. Firstly, for $e_2 =$ 0.1, we found good examples of HZ test particles that experience nodal librations throughout a full N-body experiment of 10 Myr for $e_1$ ranging between 0.4 and 0.7 (row 2 of Fig.~\ref{fig:multiplotprueba}). For $e_2$ = 0.3, our results show that nodal librations of HZ test particles survive an entire simulation for $e_1$ = 0.2, while for $e_2$ = 0.4 the survival trend of those nodal librations extends to $e_1$ values of 0.1 and 0.2 (row 3 of Fig.~\ref{fig:multiplotprueba}). For $e_2$ of 0.3 and 0.4 and $e_1$ ranging from 0.3 to 0.8, the HZ test particle’s eccentricity $e_2$ has a chaotic behavior and the temporal evolution of the ascending node longitude $\Omega_2$ frequently switches between libration and circulation throughout 10 Myr. Furthermore, we also found cases where the value of $e_2$ increases significantly, which leads to the particle being ejected from the system. We remark that all HZ particles that experience nodal librations throughout a full N-body simulation for $e_2 = 0.1, 0.3$ and 0.4 evolve on purely retrograde orbits, which is consistent with the analytical theory of this research. For $e_2$ = 0.5, we could not find HZ test particles with nodal librations capable of surviving a 10~Myr full N-body experiment for any value of $e_1$ between 0.1 and 0.8. In these cases, the value of $e_2$ changes chaotically, so that the temporal evolution of $\Omega_2$ frequently switches between libration and circulation throughout 10 Myr or else the particle ends up being ejected from the system. We highlight that these behaviors are found with the same frequency in the set of numerical simulations developed for $e_2$ = 0.5. We found similar behaviors in the HZ test particles associated with N-body simulations that assume a value of $e_2$ = 0.6. However, surprisingly, we found very good examples of HZ test particles whose nodal librations survive a full N-body experiment of 10 Myr for $e_2$ = 0.6, $e_1$ between 0.5 and 0.7, and $i_2$ ranging from  100$^{\circ}$ to 120$^{\circ}$. While all these particles should evolve on purely retrograde orbits according to our analytical criteria, we find that those with values of $i_2$ around 110$^{\circ}$ are preferably associated with such a nodal libration regime. This result is in agreement with the analytical criteria discussed in Sect. 3.1, which indicate that a HZ test particle with such orbital parameters should show librations of $i_2$ and $\Omega_2$ of very small amplitude. This behavior can be observed in the example illustrated in the row 1 of Fig.~\ref{fig:fig8_multiplot_3x3_PR}.

For an inner perturber of 3 M$_{\textrm{Jup}}$ and 5 M$_{\textrm{Jup}}$ we perform a set of N-body simulations for values of $e_2$ of 0.1, 0.3, 0.5 and 0.7. For $e_2 =$ 0.1, our numerical experiments adopt $e_1$ values between 0.1 and 0.6 and between 0.1 and 0.7 for 3 M$_{\textrm{Jup}}$ and 5 M$_{\textrm{Jup}}$, respectively. In these scenarios of work, our general results show very good examples of HZ test particle whose nodal librations survive a full N-body simulation of 10 Myr, except for an inner perturber of 3 M$_{\textrm{Jup}}$ and $e_1 =$ 0.1. Furthermore, there is a good agreement between our numerical simulations and the analytical treatment for values of $e_2$ of 0.3 and 0.5, and $e_1$ between 0.3 and 0.9. For $e_1$ of 0.1 and 0.2, the level of agreement significantly decreases, being particularly null for 3 M$_{\textrm{Jup}}$ and $e_2 = $ 0.5. Under these conditions, a high percentage of HZ test particles undergo a chaotic evolution of its eccentricity leading to nodal librations that do not survive a full N-body experiment of 10 Myr. Moreover, on the one hand, our numerical experiments show nodal librations with orbital flips for values of $e_1$, $e_2$, and $i_2$ that are in agreement with the secular treatment results for each inner perturber assumed in these scenarios of study. Examples of HZ particles that experience nodal librations with orbital flips for $m_1 =$ 3 M$_{\text{Jup}}$ and 5 M$_{\text{Jup}}$ are shown in the row 4 and 5 of Fig.~\ref{fig:multiplotprueba}, respectively. On the other hand, we observe that nodal librations correlated with purely retrograde orbits are more difficult to obtain for the space of parameters indicated by the secular treatment. 
For $m_1 =$ 3 M$_{\textrm{Jup}}$, the agreement between the N-body simulations and the secular treatment concerning nodal librations with purely retrograde orbits is good for $e_2 =$ 0.1 with $e_1$ between 0.2 and 0.6. For $e_2 =$ 0.3, such a good agreement is observed for $e_1$ between 0.7 and 0.9 and it is slightly less significant for $e_1$ ranging from 0.3 to 0.6. Finally, for $e_2 =$ 0.5, the mentioned agreement is only found for $e_1$ = 0.9. For $m_1 =$ 5 M$_{\textrm{Jup}}$, the general result shows that a good agreement between the N-body experiments and the analytical criteria concerning nodal librations with purely retrograde orbits occurs in a space of parameters that is somewhat more restrictive than that presented for $m_1 =$ 3 M$_{\textrm{Jup}}$. For $e_2$ = 0.1, the mentioned agreement is observed for $e_1$ between 0.1 and 0.7, but the fraction of N-body simulations consistent with the analytical criteria is less than that obtained for 3 M$_{\textrm{Jup}}$ and the same value of $e_2$. For $e_2$ = 0.3, such an agreement is only associated with values of $e_1$ of 0.2, 0.8, and 0.9. Finally, for $e_2$ = 0.5, the consistency between the numerical result and the secular theory concerning nodal librations with purely retrograde orbits is only found for $e_1$ of 0.1 and 0.9. A general analysis of these numerical results shows a very good consistency with that derived from the secular quadrupolar model discussed in Sect. 3.2, which indicate that the more massive the inner perturber and the greater the value of $e_2$, the smaller the nodal libration region associated with purely retrograde orbits in the ($e_1$, $i_2$) plane. A particular example of HZ particles that evolve on nodal libration trajectories with purely retrograde orbits for $m_1 =$ 3 M$_{\text{Jup}}$ and 5 M$_{\text{Jup}}$ can be observed in the row 2 and 3 of Fig.~\ref{fig:fig8_multiplot_3x3_PR}, respectively.

Finally, for $e_2 =$ 0.7, we found good examples of HZ test particles with nodal librations that survive an entire N-body simulation for an inner perturber of 3 M$_{\text{Jup}}$ and 5 M$_{\text{Jup}}$, and values of $e_1$ between 0.1 and 0.8. However, the level of agreement between our N-body experiments and the analytical treatment significantly decreases in comparison with that observed for values of $e_2$ of 0.3 and 0.5.


\section{Discussion and Conclusions}

In the present research, we study the role of the GR in the dynamical properties of outer test particles in the elliptical restricted three-body problem. In particular, we analyze the nodal librations of massless particles located at the HZ that evolve under the effects of an inner and eccentric perturber around a solar-mass star.  
First, we obtain analytical results making use of the integral of motion proposed by \citet{Zanardi2018} and \citet{Zanardi2023}, which is derived on the basis of a secular hamiltonian expanded up to the quadrupole level of the approximation. From this, we analyze the sensitivity of the nodal libration region to the eccentricity $e_2$ and inclination $i_2$ of the HZ test particle as well as to the eccentricity $e_1$ and the mass $m_1$ of the inner perturber. In this line of analysis, we find that nodal librations of a HZ test particle are possible for any value of $m_1$ and $e_1$ by adopting suitable $e_2$ and $i_2$. In fact, for a given $e_1$, the greater the $m_1$ value, the smaller the $e_2$ value above which nodal librations can be experienced. Following this correlation, we find that an inner perturber more massive than 0.84 M$_{\textrm{Jup}}$ allows the HZ test particle evolves on a nodal libration trajectory for any value of $e_2$ and an appropriate combination of $e_1$ and $i_2$. For a given $m_1$, we show that the greater the $e_2$ value, the smaller the minimum of the extreme inclination $i_2$ and the greater the maximum of $e_1$ associated with the nodal libration region. 

Our research also shows that a HZ test particle can experience nodal librations correlated to both orbital flips and purely retrograde orbits for any value of $m_1$ and $e_1$ and suitable $e_2$ and $i_2$. Our results indicate that the greater the $m_1$ value, the smaller the $e_2$ value above which nodal librations with orbital flips are possible for a given $e_1$. From this, we show that a HZ test particle perturbed by an inner super-Jupiter more massive than 1.68 M$_{\textrm{Jup}}$ can evolve on a nodal libration trajectory with orbital flip for any $e_2$ and a suitable set of values $e_1$ and $i_2$. For a given $m_1$, the greater the $e_2$ value, the greater the range of $i_2$ that leads to nodal librations with orbital flips of the HZ test particle for a given $e_1$.


The development of N-body simulations has allowed us test the robustness of the analytical criteria that lead to nodal librations of the outer test particle under GR effects. On the one hand, our results show a very good agreement between the N-body experiments and the analytical criteria derived from a secular and quadrupole theory that lead to nodal librations of a HZ test particle for a wide range of orbital parameters when an inner super-Jupiter is considered. On the other hand, when a Saturn- or Jupiter-like inner perturber is assumed, the consistency between the N-body simulations and the analytical prescriptions concerning nodal librations is just limited to a small range of values of ($e_1$, $e_2$, $i_2$). Finally, nodal librations of a HZ test particle do not survive a full N-body experiment for values of ($e_1$, $e_2$, $i_2$) determined by the analytical theory when an inner perturber less massive than Neptune is considered. According to this, the limitations of our model based on a secular and quadrupolar Hamiltonian reveal some disagreements between the analytical criteria that lead to the production and survival of nodal librations of a HZ test particle and the results derived from N-body experiments. Such as we described in Sect. 3.3, these inconsistencies occur at high $e_2$ values for each $m_1$ analyzed in the present research, and for a well-defined set ($e_1$, $e_2$, $i_2$) associated with low and moderate values of $e_2$ for an inner perturber with $m_1 \geq$ 1 M$_{\textrm{Jup}}$. The deviations observed between the analytical criteria that lead to nodal librations and the N-body simulations are due to the absence of non-secular and higher order secular terms in our model, which should play a primary role in the dynamical evolution of the particles associated with the particular cases mentioned above.


In the present research, we consider that the pericenter precession of the inner perturber is due solely to general relativity. We are aware that other effects such as tides and rotation-induced flattening  also cause a pericenter precession \citep{Sterne1939}, which could modify the analytical criteria derived in the present study associated with nodal librations of the test particle. The role of those effects over the dynamics of outer test particles in the elliptical restricted three-body problem will be the focus of study of a forthcoming paper.  

The results derived in this study can be used to study the dynamics and stability of potential objects located at the HZ of systems associated with the observational sample, which host a planet with a semimajor axis around 0.1 au orbiting around a single stellar component of solar mass. To date, 33 exoplanets have been detected and confirmed with known eccentricity, which are associated with single-planet systems orbiting a central star, whose mass ranges between 0.8 M$_{\odot}$ and 1.2 M$_{\odot}$. Those planets have an individual mass between 4.7 M$_{\oplus}$ and 10.1 M$_{\textrm{Jup}}$ and a semimajor axis ranging from 0.08 au to 0.12 au. In particular, the system around the star HAT-P-15 of 1.01 M$_{\odot}$ hosts a gaseous giant of 1.946 M$_{\textrm{Jup}}$ with a semimajor axis and an eccentricity of 0.0964 au and 0.19, respectively \citep{Kovacs2010}. The physical and orbital properties of this system make it an excellent laboratory to study the dynamics of potential objects at the HZ from the prescriptions described in the present study.

This research has allowed us to develop a detailed study that combines analytical criteria and N-body numerical experiments concerning the role of the GR in the dynamics of outer test particles in the framework of the elliptical restricted three-body problem. The application of this study to real systems will lead us to a better understanding of the stability of potential objects in the HZ, allowing a more precise and detailed description of the dynamic properties in such a peculiar region of the system.


\section*{Acknowledgements}

This work was partially financed by Agencia Nacional de Promoción de la Investigación, el Desarrollo Tecnológico y la Innovación, Argentina, through PICT 2019-2312, and Universidad Nacional de La Plata, Argentina, through the PID G172. Moreover, the authors acknowledge the partial financial support by Facultad de Ciencias Astronómicas y Geofísicas de la Universidad Nacional de La Plata, and Instituto de Astrofísica de La Plata, for extensive use of their computing facilities.

\section*{Data Availability}

All N-body simulations presented in the present manuscript will be made available
upon reasonable request to
the corresponding author.
 



\bibliographystyle{mnras}
\bibliography{Paper} 








\bsp	
\label{lastpage}
\end{document}